# RAPID RECONSTRUCTION OF 3D STRUCTURE OF FIBROUS MEDIA

*V. Berejnov*[1], *D. Sinton*[1], *and N. Djilali*[1]

[1]Department of Mechanical Engineering and Institute for Integrated Energy Systems,
University of Victoria, Victoria, V8W 3P6, Canada

**ABSTRACT**

Characterization of transport properties of porous media is increasingly relying on computational methods that require reconstruction of the media structure. We present a simple method of constructing the 3D surface of fibrous porous media – the gas diffusion layer (GDL) used as the porous electrode in PEM fuel cells. The method is based on extending the depth-of-field on the whole attainable thickness of the GDL. A series of images of the GDL sample is recorded by the sequential movement of the sample with respect to the microscope focus. Different layers of the surface of the sample appear in focus in the different images in the series. The indexed series of the in-focus portions of the sample surface is combined into one sharp 2D image and interpolated into the 3D surface representing the surface of an original GDL sample. The method uses a conventional upright stage microscope that is operated manually, the inexpensive Helicon Focus software, and the open source MeshLab software. The accuracy of the reconstruction of the image features was found to be ~ 1% for probing depth of ~ 100 µm with the corresponding image acquisition time being ~ 10 min. Applying this method we found that: *i*) this technique is able to reproduce 3D-projection of the GDL structure for a depth of ~ 50-100 µm, *ii*) the depth of the deepest channel in the GDL can be as much as ~ 100-160 µm, for GDL samples with a thickness ~ 200 µm.

**INTRODUCTION**

The transport properties of the porous material are essentially defined by the structure of the network of the interconnected voids. The modern computational methods can simulate the geometrical structure of the porous material and provide the boundary conditions to compute flow and transport of fluids. However, because of the complexity of the structure of the porous network there is always a problem of validation of the simulated results. The core part of this validation is a comparison of the simulated structure and the structure retrieved form the real porous medium. The latter requires an acquisition of the structural information of porous network imbedded in the real 3D space and reconstruction of this information into the 3D object representing the probed porous sample.

To date there are several attempts (Fluckiger, et al. 2008, Koido, et al. 2008, Inoue, et al. 2008, Fraunhofer ITWM 2009, Schulz, et al. 2007, Izzo, et al. 2008, Becker, et al. 2008) illustrating the sufficient progress in the area of computational simulation particularly related to the fibrous based porous materials. In addition, it was also demonstrated that the structural information of this kind of porous media could be obtained by using the X-ray tomography method (Koido, et al. 2008, Fraunhofer ITWM 2009, Izzo, et al. 2008, Becker, et al. 2008). However, in spite of its many advantages, the X-ray tomography requires a specific facility and substantial investing of money and time. As a result, it could be problematic to spread this technique over the wide range of labs of the fuel cell community as well as apply it to the system controlling the quality of the carbon papers in the on-line process for the fuel cell manufactures.

We developed an inexpensive and rapid in processing technique allowing retrieving the 3D structural information from a sample of the fibrous material. The technique employs the common, manually operated upright optical microscope Leica DM LM and a combination of the image processing software: Helicon Focus (HF) and MeshLab (ML). We found this approach to be appropriate for collecting the profile of the irregular rough structures of a surface of carbon paper and even for “looking” far beyond to the porous paper surface. We believe this method will benefit both the laboratory practice and the material control in mass production.

**EXPERIMENTAL AND MATERIALS**

In our experiments we used a sample of the metal mesh for calibration and the fibrous sheet of the carbon filaments (a carbon paper) for demonstration. The metal mesh was fabricated from the copper wires with 100x100 meshes per inch and the wire diameter of 0.0114 cm (0.0045’’). The porous material was the Toray carbon paper B-2/060/40WP from E-TEK Company, USA, the thickness of the sheet of this material was ~ 200 µm and the diameter of the carbon filament was about 8 µm.

The samples 2.5x2.5 cm of the copper mesh or the carbon paper were mounted horizontally on the table of the Leica DM LM microscope, used in the inverted mode. Figure 1. The resolution of positioning in the X-Y plane is 0.1 mm. The microscope focal wheel is calibrated on micron scale allowing setting the position of the Z-axis in 1 or 4 µm interval with 0.5 µm of accuracy. We used the mercury lamp (4) for the light source. Four different objectives (2) were tested: 5x/0.15, 10x/0.30, 20x/0.40, and 50x/0.50, where the first number is a magnification and the second one is a numerical aperture, see Table 1. The 8-bit images were captured with the black/white camera Retiga 1300i and preprocessed with the software QCapture Pro 6.0 from QCapture Inc.

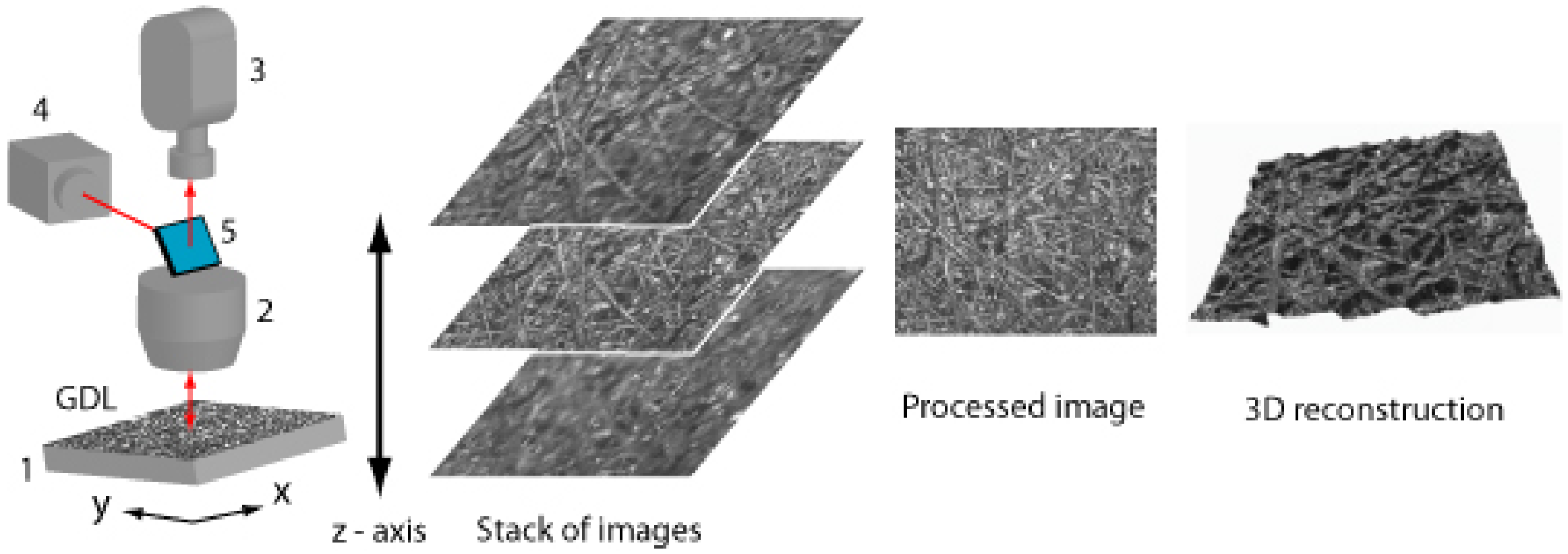

Figure 1. The sketch of the optical station and the image processing sequence. The stack of *n* images was collected, processed and reconstructed with the HF and ML software, respectively. The number denotes: (1) – microscope table movable in the X-Y-Z directions, (2) – objective, (3) – video camera, (4) – light source, and (5) – light beam divider.

## RECONSTRUCTION METHOD

When the profile of the micro-sample covers more than the attainable depth of focus, the part of the sample surface that is outside of the focus looks blurry, and thus cannot be measured and analyzed. The thickness, *d*, of an imaginary layer containing the in-focus portion of the sample surface that looks sharp is called the "depth of focus" (of the "depth of field") and abbreviated as DOF. The DOF is a function of the lenses in the objective (Boutry 1962, Bracey 1960):

$$d \approx \frac{\lambda}{Na^2},$$

where $\lambda$ is a wavelength of light that can be taken for an estimate as ~ 0.65 µm, and *Na* is a numerical aperture of the objective. Table 1 provides the values of DOF for the objectives tested in the paper.

Table 1. Parameters of the objectives

| Objective | *Ma* | *Na* | *d*, µm |
|---|---|---|---|
| HC PL Fluotar | 5x | 0.15 | 29 |
| HC PL Fluotar | 10x | 0.30 | 7.2 |
| N Plan | 20x | 0.40 | 4.1 |
| N Plan | 50x | 0.50 | 2.6 |

*Ma*, *Na*, and *d* are the objective magnification, numerical aperture, and depth of focus, respectively

Generally, the method consists of sequential taking of the multiple images (optical slices) corresponding to the different positions of the focus. The next, processing part requires the software Helicon Focus. HF selects a portion of the sample surface that is in focus from each optical slice, Figure 1. Using these portions it constructs a new, synthetic 2D image of the original surface that is sharp everywhere combining those in-focus portions into a single image. By this procedure HF extends the depth of field, *D*, over the whole attainable surface of the sample:

$$D \approx d\ n,$$

where *n* is a number of the stack images having non-overlapped elementary DOF, *d*. Next, since every in-focus portion is indexed by the appropriate slice, and the distance between slices is known, HF interpolates those assembled portions into a single 3D surface. This 3D surface represents an approximation of the surface of the original object. The details of the above technique could be found in (Berejnov 2009).

## CALIBRATION

The linear scale in the X-Y plane was calibrated with the 2 mm optical calibration standard, from the Leica Company, Germany, having the 2 mm grid scale and 10 µm tick's resolution. Recording this calibration standard with the different objectives provides the series of the X-Y scaling factors we used for the X-Y in-plane image processing. Calibration in the Z axis is less obvious because it includes multiple stages of the image processing with HF and ML software. The general idea here is to measure and compare heights and patterns of the test objects in two ways: directly and with the help of the 3D reconstruction techniques.

*Metal mesh*: The sample of the copper mesh was mounted horizontally on the microscope stage and a series of the mesh images was collected with the 10x/0.30 objective. Next, this image stack was processed with HF. As a result the 2D image of the extended field of depth and the 3D surface of the mesh sample were reconstructed, Figure 2 (a) and (b).

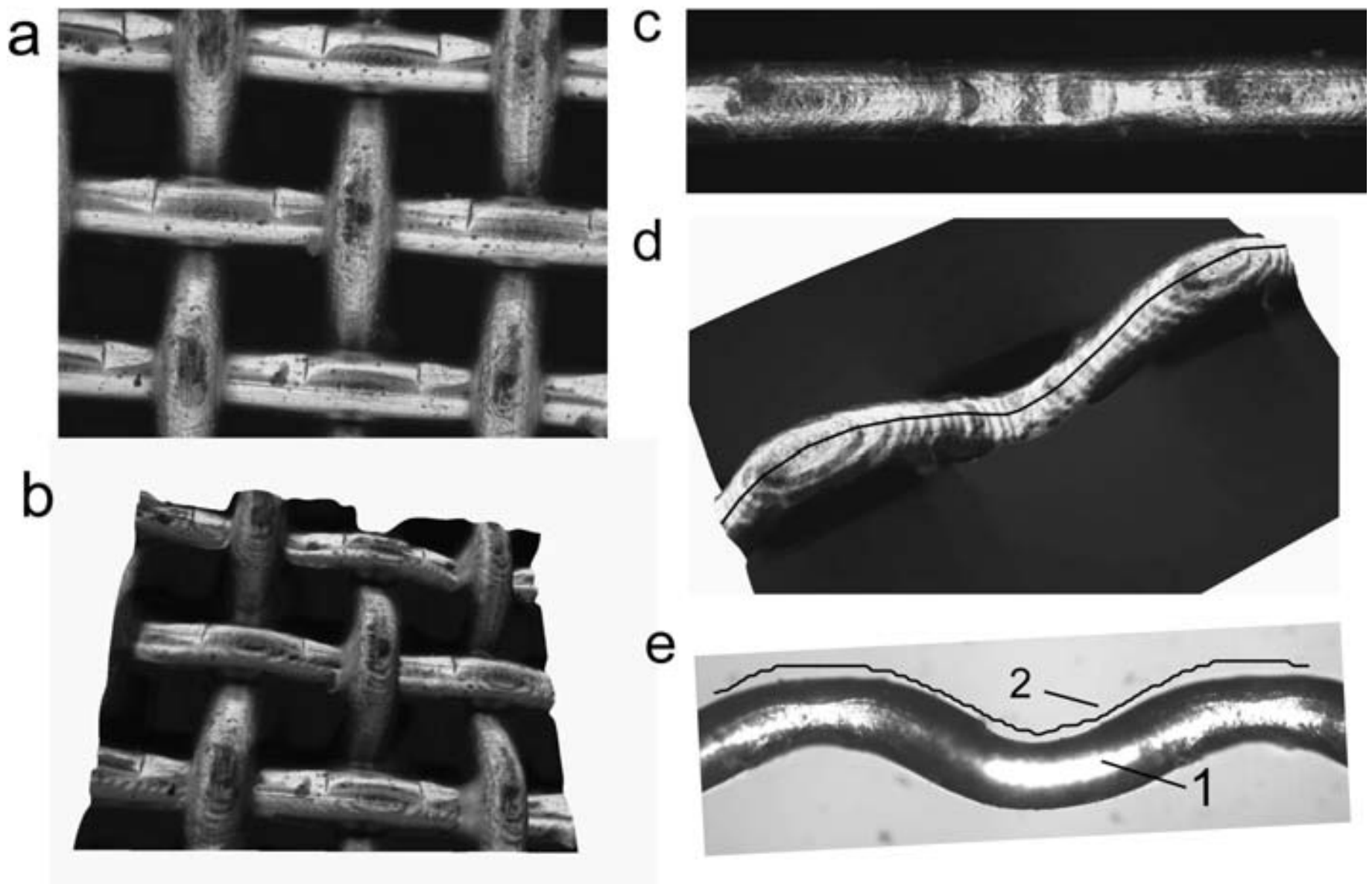

Figure 2. Reconstruction of the 3D surface of the copper mesh (a) and (b). Using the single element of the mesh (c), (d), and (e) for calibrating the scale of the Z-axis. The panels (a) and (c) represent the top views of the mesh and wire after the image processing, respectively. The panels (b) and (c) show the 3D reconstructed profiles, where the black curve on (d) depicts the cross section taken from the 3D-surface data and plotted on (e). The panel (e) presents a side view (1) of the segment shown on (c) and the profile data (2) taken from the image (d).

Then a single wire of mesh was released non-deformed from the mesh network. This single wire was positioned on the microscope stage in order to provide in the field of view, FOV, the maximal depth for the wire pitches. The image stack of this mesh element was collected for multiple focuses and processed with the HF program; Figure 2 (c) and (d). The whole in-focus image, Figure 2 (c), was obtained and the 3D surface of the mesh element was formatted as the OBJ file(Berejnov 2009).

The OBJ data was processed with the MeshLab software and the surface of the wire was rendered in the 3D space. Then the 3D surface was vertically sectioned along the length of the wire, (see the black curve in Figure 2 (d)) and the numerical profile of this section was retrieved and saved using the ML tools.

The number of the steps on the profile representing the number of the images included for the HF analysis in the stack is 14. Every step corresponds to the vertical lift of the microscope stage on 8 µm. Thus, the gap for the black curve in Figure 2 (d) corresponding to the pitch of the mesh sample is 112 µm. The direct measurement of the same gap using the image of the mesh wire turned on 90 degrees gives 114 µm, Figure 2 (e). The difference between the direct measurement and the measurement including the consecutive application of the Helicon Focus and MeshLab software is ~ 2.6%.

## RESULTS

The sample of carbon paper was mounted on the microscope stage horizontally. Then the series of the multi-focused images of the given view area were taken by varying the position of the stage along the Z axis within the 4 µm interval, *h*. Four different objectives were tested: 5x/0.15, 10x/0.30, 20x/0.40, and 50x/0.50. The stacks of the images indexed by the Z coordinate were processed with the HF software and 3D surfaces were created for every image stack, Figures 3 and 4.

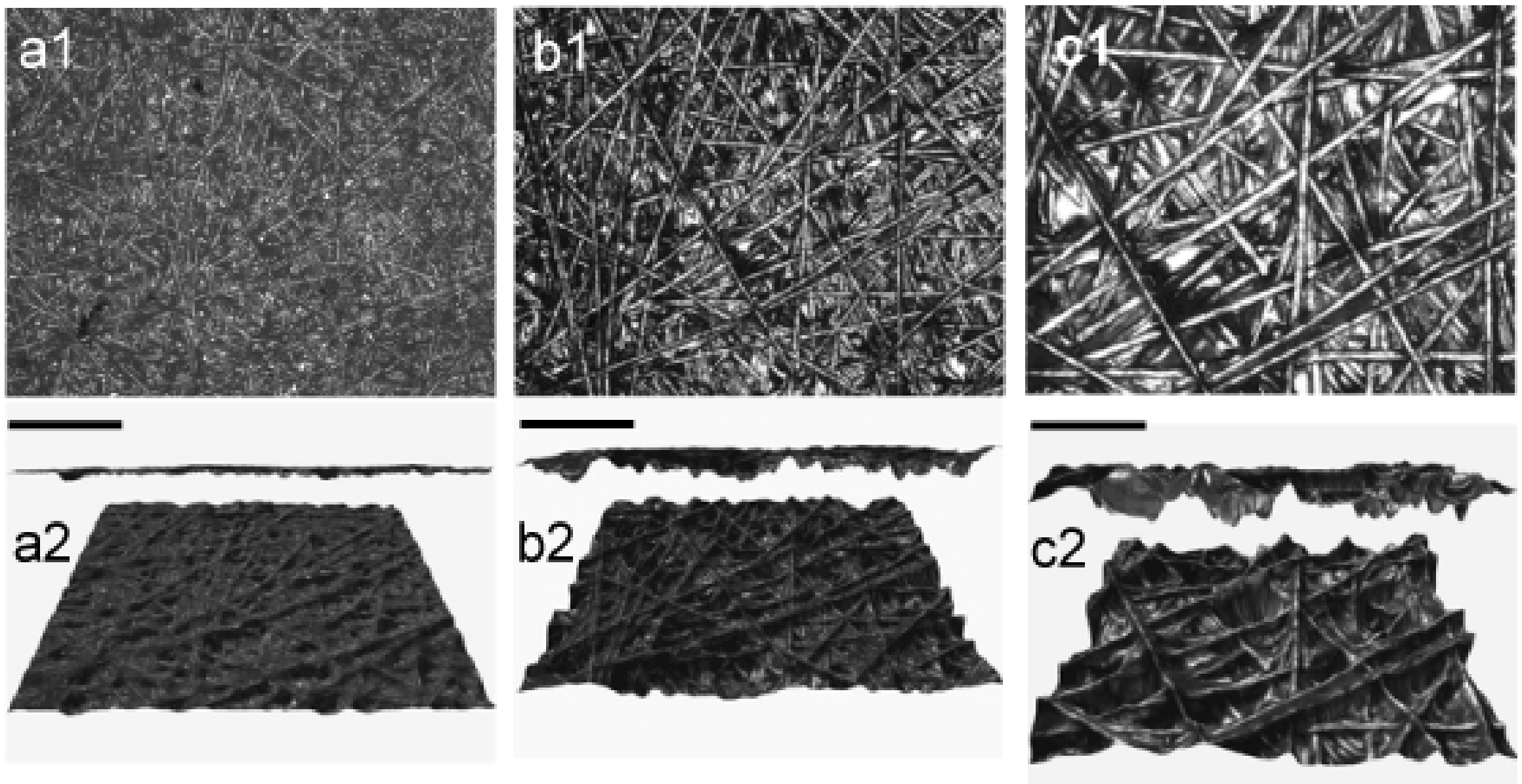

Figure 3. The processed and reconstructed images indexed 1 and 2 respectively, for three different objectives and the same view area. (a) is 5x/0.15 objective, the bar is 400 µm; (b) is 10x/0.30 objective, the bar is 200 µm; and (c) is 20x/0.40 objective, the bar is 100 µm. The 3D images are presented by the side and the perspective views of the carbon paper sample

In Figure 3 we present three tests with three different objectives acquired and processed the same area of the sample for different FOV. The values of DOF for a single optical slice, *d*, for those objectives are presented in Table 1. We can see a simple fact that if $d >> h$ (objective 5x/0.15), then the current method is able represent the very average 3D structure of the sample. In case if the interval of the stage lifting, *h*, is of order of *d* (objective 10x/30), then the current method can indeed reproduce the features of order of the size of the carbon fiber that is ~ 8 µm. For this condition the acquisition area is of order of 1 $mm^2$ that is sufficient for statistical analysis of the sample structure since the area of the features is of order of 10-100 $µm^2$. The best resolution could be achieved when *d* is equal or less than *h* (objectives 20x/0.40 and 50x/0.50, see Figure 4).

Applying the high magnification objectives provides the best quality of the 3D reconstruction of the surface structure having FOV of 0.5 $mm^2$ and 0.2 $mm^2$, respectively. This aspect makes the highest magnification objective 50x/0.50 to be very convenient for inspecting the local features of the carbon paper sample, while the objectives 20x/0.40 can be considered to be good enough for acquiring the statistics of the surface structure.

**CONCLUSION**

Using the simple manually operated microscope and an inexpensive software we developed a method for rapid constructing the 3D surface of the fibrous porous media. The method consists of three major steps: *i*) acquiring the series of images of a sample with the sequential movement of the microscope stage, *ii*) applying the Helicon Focus software for the image stack and generating the synthetic sharp image and the 3D surface of the sample, and *iii*) analyzing the 3D sample surface with the MeshLab software.

We found this method to be very convenient for inspecting the microstructured surface of the carbon paper at the different linear scales. We were able to acquire both the large scale (~1 mm) structures suitable for statistical analysis of the "whole" sample as well as the small scale (~50 µm) features useful for characterization the "local" properties of the inspected material.

**NOMENCLATURE**

| | |
|---|---|
| *d* | Depth of field (DOF) |
| *D* | Extended depth of field |
| FOV | Field of view |
| $\lambda$ | Wavelength of light |
| HF | Helicon Focus |
| ML | MeshLab |
| *Na* | Numerical aperture |
| *Ma* | Magnification |
| *h* | Moving interval of the microscope stage |
| GDL | Gas diffusion layer |
| PEM | Polyelectrolyte membrane fuel cell |

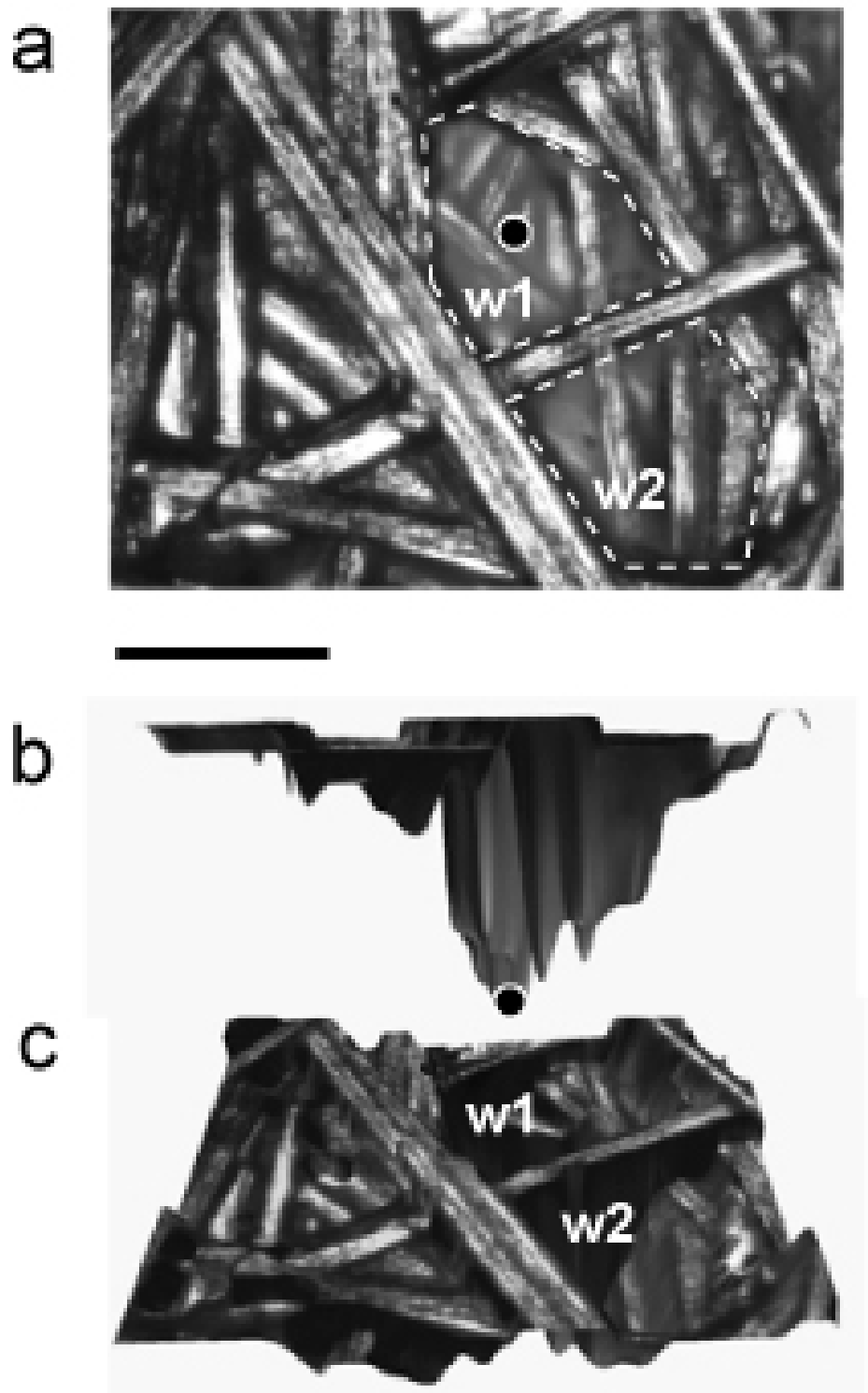

Figure 4. The processed (a) and the 3D reconstructed images (b) and (c) for the side and for the perspective views, respectively; the objective is 50x/0.50. The bar is 50 µm. There are two wells w1 and w2, the black point depicts the deepest depth of 120 µm that is located in the well w1.